\documentclass{ws-mpla-hep}
\usepackage{graphicx}
    \usepackage{amssymb}
     \usepackage{pifont}
\usepackage[figuresright]{rotating}
 \usepackage{bm}

\newcommand{\Ds}{\displaystyle}
\usepackage{color}

  \newcommand{\BluTn}[1]{\textcolor{blue}{#1}}
   \newcommand{\RedTn}[1]{\textcolor{red}{#1}}

\begin{document}
\def\preprint{RUB-TPII-12/09}
 \markboth{S.~V.~Mikhailov, N.~G.~Stefanis}
{Pion transition form factor at the two-loop level vis-\`a-vis
       experimental data}

\catchline{}{}{}{}{}
\title{
Pion transition form factor at the two-loop level vis-\`a-vis
       experimental data}

\author{\footnotesize S.~V.~Mikhailov\footnote{
 Talk presented at Workshop
 ``Recent Advances in Perturbative QCD and Hadronic Physics'',
 20--25 July 2009, ECT*, Trento (Italy),
 in Honor of Prof. Anatoly Efremov's 75th Birthday Celebration.}}
\address{Bogoliubov Laboratory of Theoretical
         Physics, JINR, 141980 Dubna, Russia \\
         mikhs@theor.jinr.ru}

\author{\footnotesize N.~G.~Stefanis\footnote{\footnotesize{Also at
        Bogoliubov Laboratory of Theoretical Physics,
        JINR, 141980 Dubna, Russia.}}}
\address{Institut f\"{u}r Theoretische Physik II,
         Ruhr-Universit\"{a}t Bochum, D-44780 Bochum, Germany \\
         stefanis@tp2.ruhr-uni-bochum.de}
\maketitle

\pub{Received (Day Month Year)}{Revised (Day Month Year)}

\begin{abstract}
We use light-cone QCD sum rules to calculate the pion-photon transition
form factor, taking into account radiative corrections up to the
next-to-next-to-leading order of perturbation theory.
We compare the obtained predictions with all available experimental
data from the CELLO, CLEO, and the BaBar Collaborations.
We point out that the BaBar data are incompatible with the convolution
scheme of QCD, on which our predictions are based, and can possibly
be explained only with a violation of the factorization theorem.
We pull together recent theoretical results and comment on their
significance.
\vspace{1pc}
\end{abstract}


\section{Introduction}
\label{sec:intro}

For many years now the production of a pion by the fusion of two
photons has attracted the attention of theorists and experimentalists.
Theoretically, the process $\gamma^*(q_1)\gamma^*(q_2)\to \pi^0(p)$
can be treated within the convolution scheme of QCD\cite{ER80,LB80}
by virtue of the factorization theorem which allows one to treat the
photon-parton interactions within perturbative QCD, while all binding
effects are separated out and absorbed into the pion distribution
amplitude (DA).
This latter ingredient has a nonperturbative origin and can, therefore,
not be computed within perturbative QCD.
One has to apply some nonperturbative approach to derive it or
reconstruct it from the data.
A widespread framework to calculate static and dynamical
nonperturbative quantities of hadrons is provided by QCD sum rules
with local\cite{CZ84} or nonlocal condensates.\cite{MR89}
This latter method was employed by us in collaboration with A.\ P.\
Bakulev (BMS)\cite{BMS01} to derive a pion DA that is able to give
good agreement with various sets of data pertaining to various pion
observables, e.g., the electromagnetic form factor\cite{BPSS04}, the
pion-photon transition form factor\cite{Ste08}, diffractive di-jets
production,\cite{BMS04kg} etc.

While the process with two off-shell photons is theoretically the
most preferable, experimentally, another kinematic situation is more
accessible, notably, when one of the photons becomes real, as probed
by the CELLO\cite{CELLO91} and the CLEO Collaborations\cite{CLEO98}.
Such a process demands more sophisticated techniques in order to take
properly into account the hadronic content of the real photon.
Indeed, first Khodjamirian,\cite{Kho99} then Schmedding and
Yakovlev,\cite{SchmYa99} used light-cone sum rules (LCSR)s\cite{BFil90}
to analyze the CLEO data, a method also applied by BMS up to the
next-to-leading-order (NLO) level of QCD perturbation
theory.\cite{BMS02}

Especially the high-precision CLEO data\cite{CLEO98} on
$F^{\gamma\gamma^{*}\pi}$ give the possibility to verify
the pion DAs quantitatively\cite{KR96,SchmYa99,BMS02,Ste08}.
It was found that the best agreement with the CLEO data
is provided by pion DAs which have
suppressed endpoints $x=0,1$, like those belonging to the
``bunch'' determined by BMS in\cite{BMS01} with the help of QCD sum
rules with nonlocal condensates.
Note that the endpoint suppression is a {\it sui generis} feature of
the nonlocality of the quark condensate and is controlled by the
vacuum quark virtuality
$\lambda_q^2\approx 0.4$~GeV${}^2$.\cite{BMS01,BMS02}
All these approaches attempt to reverse engineer the pion DA from its
(first few) moments.
In the BMS approach\cite{BMS01} the first ten moments have been
calculated, from which the corresponding Gegenbauer coefficients
$a_n$ with $n=0,2,\ldots ,10$ were determined.
It turns out that all coefficients with $n > 4$ are negligible, so
that the proposed model DA has only two coefficients: $a_2$ and $a_4$.

Recently, we\cite{MS09} extended this type of calculation to the NNLO
of QCD perturbation theory by taking into account those radiative
corrections at this order which are proportional to the
$\beta_0$-function.
More specifically, we used the hard-scattering amplitude of this order,
computed before in Ref.\ \refcite{MMP02}, in order to determine the
spectral density within the LCSR approach mentioned above.
In addition, we refined the phenomenological part of the SR by using a
Breit-Wigner ansatz to model the meson resonances.
Below, we report about the main results of this analysis and further
discuss what conclusions can be drawn by comparing the obtained
predictions with all the available experimental data.
We focus attention on the new BaBar data,\cite{BaBar09} which turn out
to be incompatible with our predictions, indicating a violation
of collinear factorization in QCD.
We perform a detailed comparison of these data with the theoretical
expectations and some proposed scenarios to explain them.

\section{Pion-photon transition form factor
$\mathbf{F^{\gamma^{*}\gamma^{*}\pi}}$ in QCD}
\label{sec:col-fac}

The transition form factor $F^{\gamma^{*}\gamma^{*}\pi}$
describes the process
$\gamma^*(q_1)\gamma^*(q_2)\to \pi^0(p)$
and is given by the following matrix element
($-q_{1}^2\equiv Q^2>0, -q_2^2\equiv q^2\geq 0$)
\begin{eqnarray}
  \int d^{4}x e^{-iq_{1}\cdot z}
  \langle
         \pi^0 (p)\mid T\{j_\mu(z) j_\nu(0)\}\mid 0
  \rangle
=
  i\epsilon_{\mu\nu\alpha\beta}
  q_{1}^{\alpha} q_{2}^{\beta}
  \cdot F^{\gamma^{*}\gamma^{*}\pi}(Q^2,q^2)\, .
 \label{eq:matrix-element}
\end{eqnarray}
Provided the photon momenta are sufficiently large
$Q^2, q^2 \gg m_\rho^2$ (where the  hadron scale is set by the
$\rho$-meson mass $m_\rho$), the pion binding effects can be
absorbed into a universal pion distribution amplitude of twist-two.
Then, one obtains the form factor in the form of a convolution by
virtue of the collinear factorization:\cite{ER80,LB80}
\begin{eqnarray}
  F^{\gamma^{*}\gamma^{*}\pi}(Q^2,q^2)
=
  T(Q^2,q^2,\mu^2_{\rm F};x)
\otimes
  \varphi^{(2)}_{\pi}(x;\mu^2_{\rm F})
 + \ O\left( Q^{-4} \right) \, .
\label{eq:convolution}
\end{eqnarray}
Here the pion DA $\varphi^{(2)}_{\pi}$ represents a
parametrization of the pion matrix element at the (low) factorization
scale $\mu^2_{\rm F}$, whereas the amplitude $T$, describing the hard
parton subprocesses, can be calculated in QCD perturbation theory:
$T=T_0+a_s\, T_1+a_s^2\, T_2\ldots$,
where $a_s = \alpha_s/(4\pi)$,
and with $O\left( Q^{-4} \right)$ denoting the twist-four contribution.
In leading order (LO) of the strong coupling and taking into account
the twist-four term explicitly, one has\cite{Kho99}
\begin{eqnarray}
   F^{\gamma^{*}\gamma^{*}\pi}(Q^2,q^2)
\! =  N_f\left[
               \int_{0}^{1} \! dx
               \frac{\varphi^{(2)}_{\pi}(x;\mu^2_{\rm F})}
               {Q^{2}x + q^{2}\bar{x}}~
               -\delta^2(\mu^2_{\rm F})\int_{0}^{1} \! dx
               \frac{\varphi^{(4)}_{\pi}(x;\mu^2_{\rm F})}
               {(Q^{2}x + q^{2}\bar{x})^2}
         \right]
&&
\label{Eq:T_0}
\end{eqnarray}
with $N_f=\frac{\sqrt{2}}{3}f_{\pi}$ and $\bar{x}\equiv 1-x$.
The pion DA of twist-two, $\varphi^{(2)}_{\pi}$, is defined by
\begin{eqnarray}
  \langle
         0|\bar{q}(z)\gamma_{\mu}\gamma_{5}\mathcal{C}(z,0) q(0)|\pi(P)
  \rangle
  \Big|_{z^2=0}~=
  ~iP_{\mu} f_\pi\int dx
  e^{ix(z\cdot p)}\varphi^{(2)}_\pi(x,\mu^2_{\rm F}) \, ,&&
\label{eq:pion-DA}
\end{eqnarray}
where
$
 \mathcal{C}(z,0)
=
 \mathcal{P}\exp \left(ig\int^z_0 A_\mu(\tau) d\tau^\mu \right)
$ is a path-ordered exponential to ensure gauge invariance.
The second term in Eq.\ (\ref{Eq:T_0}) represents the twist-four
contribution, which is becoming important for small and intermediate
values of $Q^2$.
The pion DA of twist-four, $\varphi^{(4)}_{\pi}$, is an effective
one\cite{BFil90} in the sense that it is composed from different pion
DAs of twist-four.
In our analysis it is taken in its asymptotic form.
The parameter $\delta^2$ is determined from the matrix element
$
 \langle
        \pi(P)|g_s\bar{d}\tilde{G}_{\alpha \mu}\gamma^{\alpha}u|0
 \rangle
=
 i\delta^2 f_\pi p_\mu
$, and is estimated\cite{BMS02} to be
$\delta^2(1$\ GeV$^2)=0.19\pm 0.02$GeV$^2$.
Estimates for the twist-four term, based on the renormalon
approach,\cite{BGG04} have been considered in the last entry of Ref.\
\refcite{BMS02}.

On the other hand, also the evolution of
$\varphi^{(2)}_\pi(x,\mu^2_{\rm F})$ with
$\mu^2_{\rm F}$
is controlled by a perturbatively calculable evolution
kernel $V$, following the
Efremov-Radyushkin-Brodsky-Lepage (ERBL)\cite{ER80,LB80}
equation
\begin{eqnarray}
  \mu^2 \frac{d}{d\mu^2}\varphi(x;\mu^2)
=
  \left(a_s\, V^{(0)}(x,y)+a_s^2\, V^{(1)}(x,y)+\ldots \right)
  \otimes
  \varphi(y;\mu^2)
&&\label{eq:ERBL} \\
  V^{(0)}
  \otimes
  \psi_n
 =
  2C_\text{F}~v(n) \cdot \psi_n;~~~~~~ \! \psi_{n}(x)
=
  6x\bar{x}~C^{(3/2)}_{n}(x-\bar{x});
&& \\
  v(n)
=
  1/(n+1)(n+2)-1/2+2(\psi(2)-\psi(n+2));~~~
  \psi(z)
=
  \frac{d}{dz}\ln(\Gamma(z)) \, .&&
\end{eqnarray}
Here, $\{\psi_{n}(x)\}$ are the Gegenbauer harmonics, which
constitute the LO eigenfunctions of the ERBL equation,
$v(n)$ being the corresponding eigenvalues.
Then, one has
\begin{eqnarray}
  \varphi^{(2)}_{\pi}(x;\mu^2)
=
  \psi_0(x) + \sum_{n=2,4,\ldots} a_{n}(\mu^2)~\psi_{n}(x)\, ,
\end{eqnarray}
where the coefficients $\{a_n \}$ evolve (in LO) with $\mu^2$
and have specific values for each pion DA model (a compilation of the
coefficients of various proposed models can be found in
Refs.\ \refcite{BMS02,Ste08}).

The radiative corrections to the hard amplitudes in NLO, encapsulated
in $T_1$, have been computed in Ref.\ \refcite{DaCh81}.
More recently, the $\beta$--part of the NNLO amplitude $T_2$,
i.e., $\beta_0 \cdot T_{2,\beta}$, was also calculated.\cite{MMP02}
It is instructive to discuss the structure of this result at
the scale $\mu^{2}_{\rm F}=\mu^{2}_{\rm R}$, especially in view of
further considerations in Sec.\ \ref{sec:LCSR}:
\begin{eqnarray}
  \beta_0 T_{2,\beta}
=
  \beta_0T_0
  \otimes
  \left[
          C_{\rm F} {\cal T}^{(2)}_{\beta}
        - C_{\rm F} {\rm L}(y)
        \cdot
        {\cal T}^{(1)}
        + {\rm L}(y)\cdot \left(V^{(1)}_{\beta}\right)_
        + -\frac{1}{2}{\rm L^2}(y)
        \cdot
        V^{(0)}
  \right]\,  ,~
\label{eq:T2beta}
\end{eqnarray}
where ${\rm L}(y)=\ln\left[(Q^2y +q^2\bar{y})/\mu^{2}_{\rm F} \right]$.
The first term, $C_{\rm F} {\cal T}^{(2)}_{\beta}$, is the
$\beta_0$-part of the NNLO coefficient function and represents, from
the computational point of view, the most cumbersome element of the
calculation in Ref.\ \refcite{MMP02}.
The next term originates from the NLO coefficient function
${\cal T}^{(1)}$ using one-loop evolution of $a_s$.
The third term appears as the $\beta_0$-part of the two-loop ERBL
evolution (see Eq.\ (\ref{eq:ERBL})), while
the last term, which is proportional to $V^{(0)}$, stems from
the combined effect of the ERBL-evolution and the one-loop evolution
of $a_s$.

\section{Light Cone Sum Rules for the process
$\mathbf{\gamma^*(Q^2)\gamma(q^2\simeq 0) \to \pi^0}$}
\label{sec:LCSR}

The transition form factor, when one of the photons becomes quasi
real ($q^2 \to 0$), has been measured by different
Collaborations.\cite{CELLO91,CLEO98,BaBar09}
However, this kinematics requires the modification of the standard
factorization formula Eq.\ (\ref{eq:convolution}) in order to take
into account the long-distance interaction, i.e., the hadronic content
of the on-shell photon.
A viable way to reach this goal is to employ the method of LCSRs,
which are based on a dispersion relation for
$F^{\gamma^{*}\gamma^{*}\pi}$ in the variable $q^2$, viz.,
\begin{equation}
  F^{\gamma^{*}\gamma^{*}\pi}\left(Q^2,q^2\right)
=
  \int_{0}^{\infty} ds
  \frac{\rho\left(Q^2,s\right)}{s+q^2} \, .
\label{eq:dis-rel}
\end{equation}
The key element in this equation is the spectral density
$
 \rho(Q^2,s)
=
 \frac{\mathbf{Im}}{\pi}
 \left[F^{\gamma^*\gamma^*\pi}(Q^2,-s)
 \right]
$
for which we make the ansatz\cite{Kho99}
$
 \rho
=
 \rho^{\rm ph}(Q^2,s) \theta(s_0-s)
 + \rho^{\rm PT}(Q^2,s) \theta(s-s_0)
$,
where  the ``physical'' (ph) spectral density $\rho^{\rm ph}$
serves to accommodate the hadronic content of the photon
(below an effective threshold $s_0$) by means of the transition
form factors $F^{\gamma^*V \pi}$ of vector mesons, notably,
$\rho$ or $\omega$,
\begin{equation}
  \rho^{\rm ph}(Q^2,s)
=
  \sqrt{2}f_\rho F^{\gamma^*V \pi}(Q^2)
  \cdot
  \delta(s-m^2_{V}) \, .
\label{eq:resonance}
\end{equation}
On the other hand, $\rho^{\rm PT}$ embodies the partonic part of the
LCSR and derives from Eq.\ (\ref{eq:convolution}) via the relation
$
 \rho^{\rm PT}(Q^2,s)
=
 \frac{\mathbf{Im}}{\pi}
 \left[F^{\gamma^*\gamma^*\pi}(Q^2,-s)
  \right]
$.
The pion-photon transition form factor
$F^{\gamma^{*}\gamma\pi}(Q^2,0)$
can be expressed in terms of $\rho^{\rm PT}$ having recourse to
quark-hadron duality in the vector channel:
\begin{eqnarray}
  F^{\gamma\gamma^*\pi}_\text{LCSR}(Q^2)
&=&
  \frac{1}{\pi}\int_{s_0}^{\infty}%
  \frac{\textbf{Im}\left(F^{\gamma^*\gamma^*\pi}(Q^2,-s)\right)}{s} ds
\nonumber \\
&&
  +  \frac{1}{\pi}\int_{0}^{s_0}%
  \frac{\textbf{Im}\left(F^{\gamma^*\gamma^*\pi}(Q^2,-s)\right)}%
  {m_\rho^2}
            e^{(m_\rho^2-s)/M^2}ds \ ,
\label{eq:LCSR}
\end{eqnarray}
with $s_0 \simeq 1.5$~GeV$^2$ and $M^2$ denoting the Borel parameter
in the interval ($0.5-0.9$)~GeV$^2$.
The LO spectral density $\rho^{(0)}$ has been determined in Ref.\
\refcite{Kho99} using for $F^{\gamma^*\gamma^*\pi}$ Eq.\
(\ref{Eq:T_0}).

Partial results for the NLO spectral density $\rho^{(1)}$ in the
leading-twist approximation have been given in Ref.\ \refcite{SchmYa99},
while the general solution
$
 \rho_n^{(1)}(Q^2,s)
 =
 \frac{\mathbf{Im}}{\pi}
                        \left[\left(T_{1}\otimes \psi_n\right)(Q^2,-s)
                        \right]
$
was obtained in Ref.\ \refcite{MS09} to read
\begin{equation}
  \rho_n^{(1)}(Q^2,s)
=
  \frac{\rho^{(1)}_n\left(x;\mu^2_{\rm F}\right)}
       {(Q^2+s)}\Bigg|_{x=\frac{Q^2}{Q^2+s}}
\end{equation}
with
\begin{eqnarray}
  \rho^{(1)}_n\left(x;\mu^2_{\rm F}\right)
= &&
  C_{\rm F} \left[
                   -3\left[1-v^{a}(n)\right]+\frac{\pi^2}{3}
                   -\ln^2\left(\frac{\bar{x}}{x}\right) + 2v(n)
                   \ln\left(\frac{\bar{x}}{x} \frac{Q^2}{\mu^2_{\rm F}}
                      \right)
            \right] \psi_n(x)
\nonumber \\
&& \!\!\!
    - C_{\rm F}\, 2\!\!\!
    \sum^n_{l=0,2,\ldots}
    \left[G_{nl}+v(n)\cdot b_{n l}\right] \psi_l(x) \, ,~
\label{eq:spec-den-NLO}
\end{eqnarray}
where $v^a(n)= 1/(n+1)(n+2)-1/2$ and $G_{nl},~b_{n l}$ are calculable
triangular matrices (see Ref.\ \refcite{MS09} for details).
Note that the spectral density $\rho^{(1)}_n$ allows one to obtain
$F^{\gamma\gamma^*\pi}_\text{LCSR}(Q^2)$
for any number of the Gegenbauer harmonics in the expansion of
$\varphi_\pi$.

\begin{figure}[hbt]
 \centerline{\includegraphics[width=0.47\textwidth]{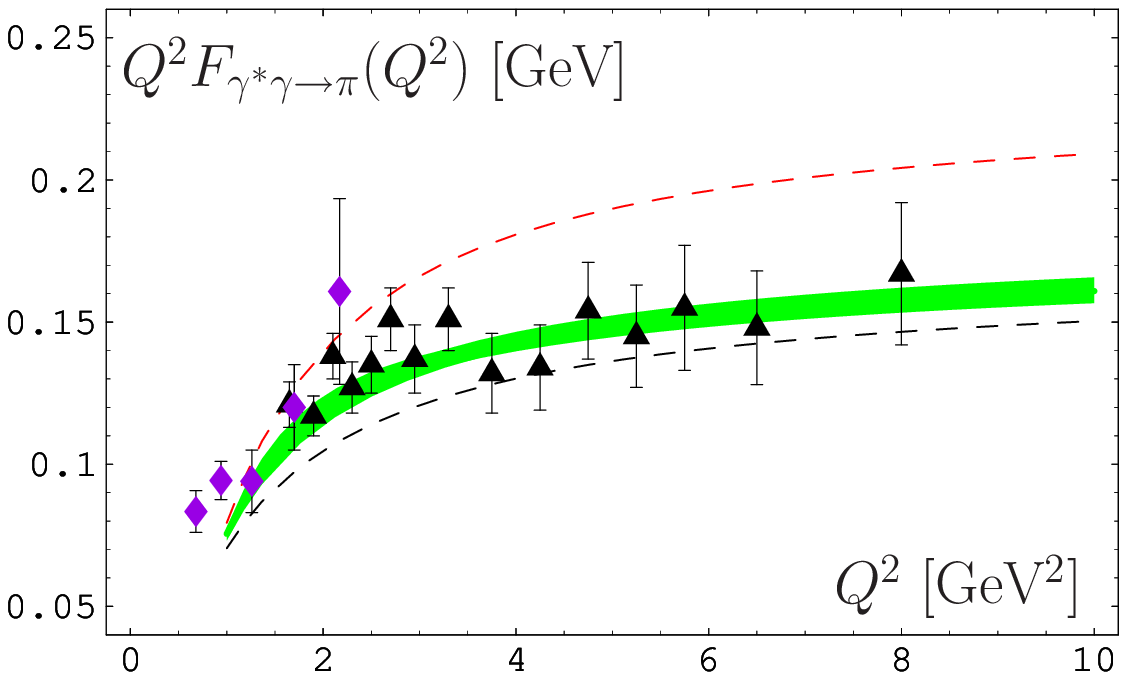}~~~~~
             \includegraphics[width=0.48\textwidth]{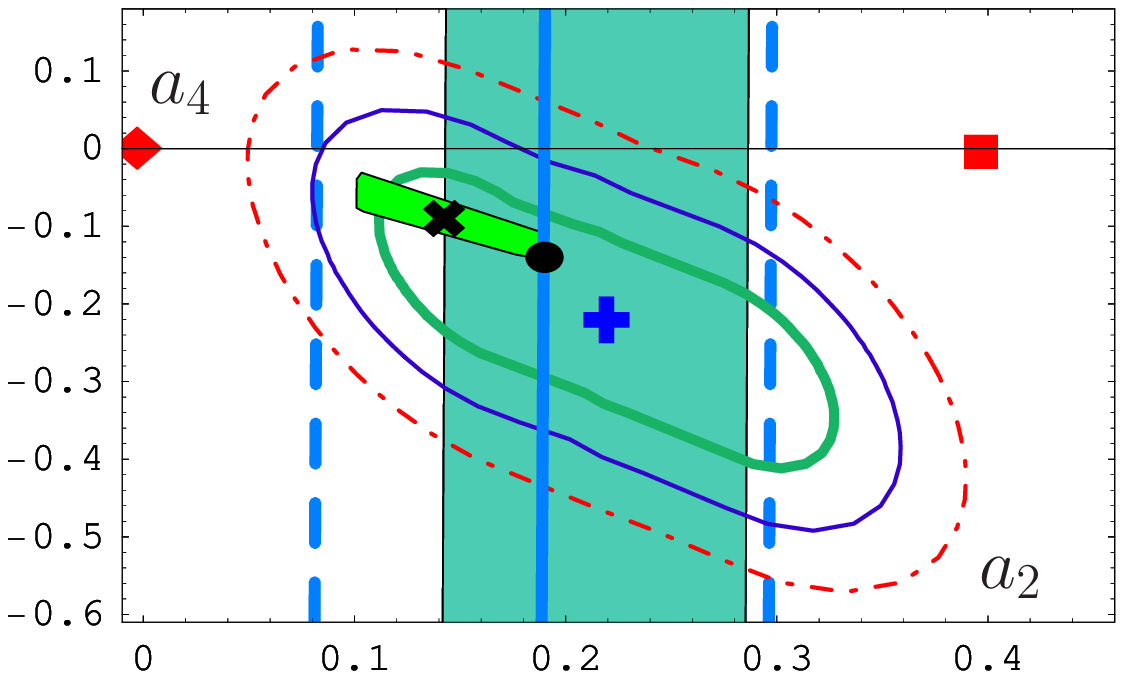}}
 \vspace*{-3mm}
 \caption{\footnotesize
\textbf{Left}: Predictions for
 $Q^2 F^{\gamma\gamma^*\pi}_\text{LCSR}(Q^2)$
 using the CZ model (upper dashed red line), the BMS ``bunch''
 (shaded green strip), and the Asy DA (low dashed black line)
 in comparison with the CELLO (diamonds) and the CLEO (triangles) data.
\textbf{Right}:
 CLEO-data constraints on
 $F^{\gamma\gamma^*\pi}_\text{LCSR}(Q^2)$
 in the ($a_2$, $a_4$) plane
 at the scale $\mu^2=(2.4$\,GeV$)^2$ in terms of error regions around
 the BMS best-fit point {\BluTn{\ding{58}}},{\protect\cite{BMS02}}
 using the following designations:
 $1\sigma$ (thick solid green line);
 $2\sigma$ (solid blue line);
 $3\sigma$ (dashed-dotted red line).
 Two recent lattice simulations{\protect\cite{Lat06,Lat07}} are
 denoted, respectively, by vertical dashed and solid lines together
 with predictions of QCD sum rules with nonlocal condensates (slanted
 green rectangle),
 {\ding{54}} --- BMS model,{\protect\cite{BMS01}}
 \RedTn{\ding{117}} --- asymptotic DA,
 \RedTn{\footnotesize \ding{110}} --- {CZ} DA.{\protect\cite{CZ84}}
}
\label{F:DAmodels}
\end{figure}

Employing this approach, predictions at the NLO for
$Q^2F^{\gamma\gamma^*\pi}_\text{LCSR}(Q^2)$ were derived,\cite{BMS02}
using a  variety of pion DAs: the asymptotic (Asy) one, the CZ model,\cite{CZ84}
and the BMS ``bunch''\cite{BMS01} (see left panel of
Fig.\ \ref{F:DAmodels}).
It was found that the radiative corrections are important, being
negative and contributing up to --17\% at low and moderate $Q^2$
values.
Let us recall the main results of this CLEO-data analysis, referring to
Refs.\ \refcite{BMS02,Ste08} for a full-fledged discussion and more
details.
The CLEO data\cite{CLEO98} were processed in terms of $\sigma$ error
ellipses and the results are displayed in the right panel of
Fig.\ \ref{F:DAmodels} around the best fit point
({\BluTn{\ding{58}}}).
In this graphics, the BMS ``bunch'' of pion DAs\cite{BMS01} is shown as
a slanted green rectangle, while the vertical dashed and solid lines
denote the estimates for $a_2$ (related to the second moment of
$\varphi_\pi$) of two recent lattice simulations in Refs.\
\refcite{Lat06,Lat07}, respectively.
Several other models, explained in Refs.\ \refcite{BMS02,Ste08}, are
also shown.
The main upshot is that wide pion DAs, like the CZ model, are excluded
at the 4$\sigma$ level, whereas the asymptotic DA, and others close to
it, are also excluded at the level of at least 3$\sigma$.
As one sees from this figure, only endpoint-suppressed pion DAs of the
BMS type are within the $1\sigma$ error ellipse of the CLEO
data\cite{CLEO98} and simultaneously in agreement with the mentioned
lattice constraints.

The inclusion of the main, i.e., $\beta_0$-proportional NNLO
contribution, in $F^{\gamma\gamma^*\pi}_\text{LCSR}$ proceeds
via the dispersion integral in (\ref{eq:LCSR}).\cite{MS09}
The technical problem is how to obtain the contributions to
$\rho^{(2,\beta)}$ from terms with various powers of the
logarithms ${\rm L}(y)$ in Eq.\ (\ref{eq:T2beta}).
The outcome of this calculation turns out to be negative,
like the NLO contribution, and about --7\% (taken together with
the effect of a more realistic Breit-Wigner (BW) ansatz for the
meson resonances in Eq.\ (\ref{eq:resonance})) at small
$Q^2\sim 2$~GeV$^2$.
The size of this suppression decreases rather fast to --2.5\% with
increasing $Q^2 \geq 6$~GeV$^2$.
\begin{figure*}[hbt]
{\includegraphics[width=0.48\textwidth]{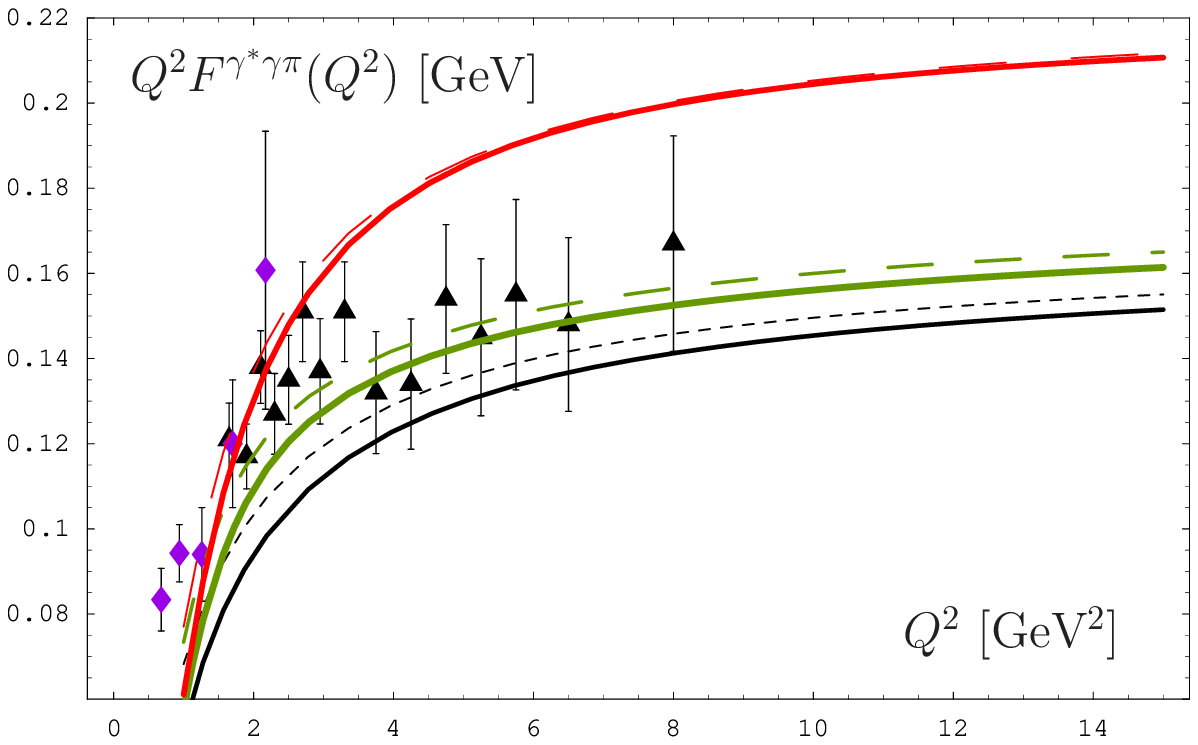}~~~~~
 \includegraphics[width=0.48\textwidth]{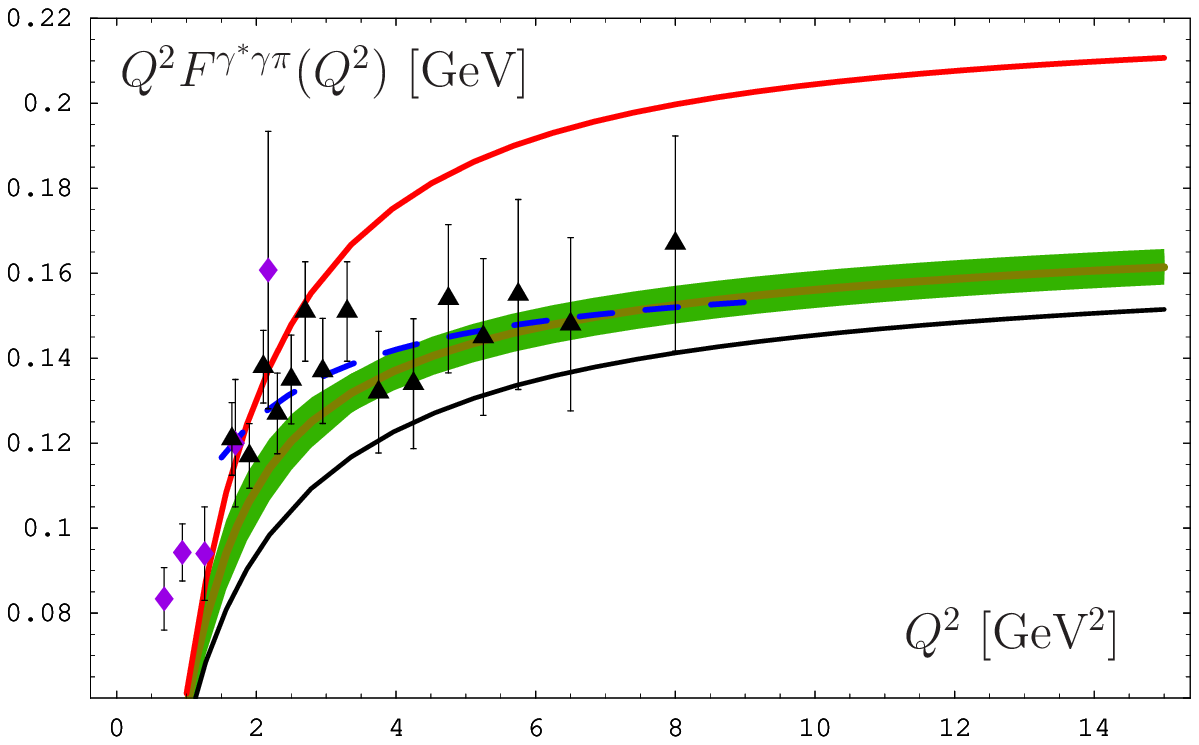}}
 \vspace*{-3mm}
\caption{\footnotesize
\textbf{Left:}
 Predictions  for
 $Q^2 F^{\gamma\gamma^*\pi}_\text{LCSR}$ with the
 NNLO$_\beta$ corrections (together with the BW resonance model)
 included (solid lines) and without them (dashed lines).
 They correspond to selected pion DAs:
 CZ model --- upper (red) line,\protect\cite{CZ84}
 BMS-model --- middle (green) line,\protect\cite{BMS01}
 and Asy DA --- lower (black) line.
 The experimental data are as in Fig.\ \ref{F:DAmodels}.
\textbf{Right:}
 The shaded (green) strip denotes the form-factor predictions derived
 for the BMS ``bunch''.\protect\cite{BMS01}
 The dashed (blue) line represents a dipole-form interpolation
 of the CLEO data.}
\label{fig:NNLOvsCLEO}
\end{figure*}
The net result is a slight suppression of the prediction for the
scaled form factor (see Fig.\ \ref{fig:NNLOvsCLEO}).

\section{Confronting NNLO LCSR results with the BaBar data}
\label{sec:NNLO-BaBar}

\begin{figure}[ht]
\centerline{\includegraphics[width=0.65\textwidth]{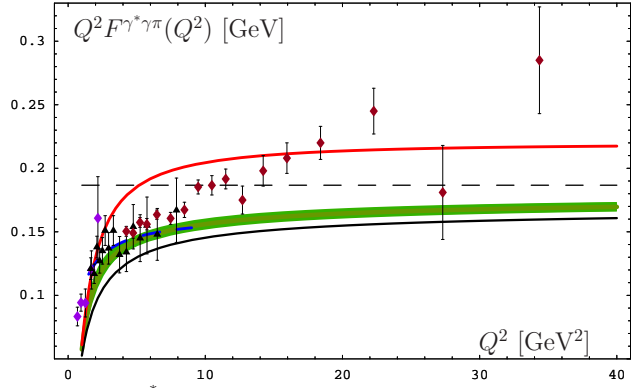}}
 \vspace*{-2mm}
   \caption{\footnotesize
 Predictions for
 $Q^2 F^{\gamma\gamma^*\pi}_\text{LCSR}(Q^2)$ calculated with
 the following pion DAs: Asy --- lower solid line, BMS ``bunch''
 --- shaded green strip, and the CZ model --- upper solid red line.
 The BaBar data\protect\cite{BaBar09} are shown as diamonds
 with error bars.
 The CELLO\protect\cite{CELLO91} and the CLEO\protect\cite{CLEO98}
 data are also shown with the designations used in Fig.\
 \ref{F:DAmodels}.
 The displayed theoretical results include the NNLO$_{\beta}$
 radiative corrections and the BW model for the meson resonances.
 The horizontal dashed line marks the asymptotic QCD
 prediction $\sqrt{2}f_\pi$.}
\label{fig:NNLO-BaBar}
\end{figure}

In the preceding section we have shown in detail that the CLEO data
are strictly incompatible with wide pion DAs and demand that the
endpoints $x=0,1$ are stronger suppressed than in the asymptotic
DA.
Surprisingly, the new data of the BaBar Collaboration\cite{BaBar09}
on the pion-photon transition form factor are in contradiction
with this behavior.
More specifically, these data, which extend from intermediate up to
high momenta in the range $4 < Q^2 < 40$~GeV$^2$, show a significant
growth with $Q^2$ for values above $\sim 10$~GeV$^2$.
Indeed, the corresponding data points lie above the asymptotic QCD
prediction $\sqrt{2}f_{\pi}$ and continue to grow with $Q^2$ up to
the highest measured momentum.
This behavior of the BaBar data is clearly in conflict with the
collinear factorization.
This in turn means that, as we argued in Ref.\ \refcite{MS09}, the
inclusion of the NNLO radiative corrections cannot reconcile the
BaBar data with perturbative QCD.
This is true for any pion DA that vanishes at the endpoints $x=0,1$
(see Fig.\ \ref{fig:NNLO-BaBar}).
From Table \ref{table:1} it becomes clear that also a wide
pion DA, like the CZ model, though it lies above the asymptotic
prediction shows exactly the same scaling behavior above
$\sim 10$~GeV$^2$ as all discussed pion DAs, which is not even
surprising because it also vanishes at the endpoints $x=0,1$.
The conclusion is that --- contrary to the statements of the BaBar
Collaboartion\cite{BaBar09} --- also the CZ DA cannot reproduce
\emph{all} BaBar data.\cite{MS09,MS09HSQCD09,Kho09}
Indeed, from Fig.\ \ref{fig:NNLO-BaBar} one sees that in the region
of the CLEO data, the CZ model fails also with respect to the BaBar
data points at the $4\sigma$ level, whereas above
$\simeq 20$~GeV${}^2$, and up to the highest measured value of $Q^2$,
$40$~GeV${}^2$, it fails again, because instead of growing with $Q^2$
it scales.
\begin{table}[ht]
 \tbl{ \footnotesize $\chi^2_{ndf}$ for the Asymptotic (Asy), BMS, and
 CZ DAs.}
{\begin{tabular}{p{0.05\textwidth} p{0.25\textwidth} p{0.20\textwidth} p{0.3\textwidth}} \hline
   Pion DA &BaBar \& CLEO data $\chi^2_{ndf}$&\hbox{BaBar all~data} $\chi^2_{ndf}$&\hbox{BaBar only highest 10 data} $\chi^2_{ndf}$ \\
   \hline
  Asy           & $11.5$   ~~~           & $19.2$      ~~             &$19.8$ ~~ \\
 BMS   & $4.4$~~~&$ 7.8$~~    & $11.9$ ~~ \\
 CZ       & $20.9$   ~& $36.0$     ~~& $6.0$  ~~ \\                                      \hline
\end{tabular} \label{table:1}
}
\end{table}

In summary, \\
(i) The combined effect of the negative ${\rm NNLO}_\beta$, i.e.,
suppressing, radiative corrections and the enhancing effect of using a
Breit-Wigner model for the vector-meson resonances finally amounts to
a moderate overall suppression of
$Q^2 F^{\gamma\gamma^*\pi}_\text{LCSR}(Q^2)$ in the range of
momentum transfer 10-40~GeV$^2$,\cite{MS09,MS09HSQCD09} probed in the
BaBar experiment. \\
(ii) The growth with $Q^2$ of the scaled form factor, measured by
the BaBar Collaboration, cannot be attributed to the hadronic content
of the real photon because this is a twist-four contribution that is
rapidly decreasing with increasing $Q^2$. \\
(iii) It is impossible to get enhancement of the form factor within
the QCD collinear factorization using pion DA models which have a
convergent projection onto the Gegenbauer harmonics and, hence, vanish
at the endpoints $x=0, 1$
(cf.\ the $\chi^2$ values in Table \ref{table:1}).
It seems (see next section) that exactly the violation of this
feature may provide an explanation of the BaBar data.
\hspace*{-5mm}

\section{BaBar data --- heuristic explanations}
\label{sec:BaBar-scenarios}

Despite the claims by the BaBar Collaboration\cite{BaBar09} that
their data are in agreement with QCD, such an explanation is outside
of reach at present.
Hence, once is forced to look for alternative explanations.
There have been several proposals to explain the anomalous behavior
of the BaBar data, among others, e.g., Refs.\
\refcite{Rad09,Pol09,Dor09,LiMi09,KoPr09,BA09}.
We restrict attention to one sort of such proposals which assumes that
the pion DA may be ``practically flat'', hence violating the collinear
factorization and entailing a (logarithmic) growth of
$Q^2 F^{\gamma^*\gamma\pi}(Q^2)$ with $Q^2$.
One has\cite{Rad09} ($\sigma^2 =0.53$~GeV$^2$)
\begin{equation}
  Q^{2}F^{\gamma^{*}\gamma\pi}
=
  \frac{\sqrt{2}f_{\pi}}{3}
  \int_{0}^{1} \frac{1}{x}
  \Ds \left[1 - {\rm e}^{ -\frac{x Q^2}{\bar{x}2 \sigma}}\right]dx \, .
\label{eq:flat-Rad}
\end{equation}
Another option\cite{Pol09} gives instead
($m\approx 0.65$~GeV, cf.\ Eq.\ (\ref{Eq:T_0}))
\begin{equation}
  Q^{2}F^{\gamma^{*}\gamma\pi}
=
  \frac{\sqrt{2}f_{\pi}}{3}
  \int_{0}^{1}
  \frac{\varphi_{\pi}(x,Q)}{x + \frac{m^2}{Q^2}}dx\,
\label{eq:flat-Pol}
\end{equation}
with $\varphi_{\pi}(x,\mu_{0})=N+(1-N)6x\bar{x}$ with
$N\approx 1.3$ and $\mu_0=0.6-0.8$~GeV.
Equation (\ref{eq:flat-Rad}) can be compared with the available
experimental data\cite{CELLO91,CLEO98,BaBar09} using for them
the phenomenological fit
($\Lambda\approx 0.9$~GeV, $b\approx-1.4$)
$$
  Q^{2}F^{\gamma^{*}\gamma\pi}
= \!
  \frac{Q^2}{2 \sqrt{2} f_{\pi} \pi^2}
  \!\left[\frac{\Lambda^2}{\Lambda^2+Q^2}+
  b \left(\!\!\frac{\Lambda^2}{\Lambda^2+Q^2} \!\!\right)^2 \right]
\label{eq:dipole}
$$
\vspace*{-1mm}
 \begin{table}[htb]
\tbl{ $\chi^2_{ndf}$ for the phenomenological fit
      and the fit with a flat pion DA, like Eq.\
      \protect(\ref{eq:flat-Rad}) --- numbers in parentheses.
      The diagonal elements give the best-fit values of
      $\chi^2_{ndf}$ which fix the corresponding line parameters.
  }
{\begin{tabular}{p{0.21\textwidth} p{0.21\textwidth} p{0.16\textwidth}}
               &  CELLO\& CLEO
               &  BaBar  \\  \hline
 \hspace*{-2mm}CELLO\& \hspace*{-2mm}CLEO
               &  $0.48~(1.22)$  & $7.8~(15.8)$      \\ \hline
 \hspace*{-2mm}BaBar$\vphantom{^\big|_|}$
               &  $10.8~(3.5)$   & $1.8~(1.8)$  \\ \hline
\end{tabular}
 \label{table:dipole-flat}
 }
  \end{table}
by means of a $\chi^2_{ndf}$ criterion, given in Table
\ref{table:dipole-flat}.
One observes --- diagonal from the BaBar (CELLO\&CLEO) entry to the
CELLO\&CLEO (BaBar) one --- that it is not possible to fit
simultaneously the CELLO/CLEO data and the BaBar data with the same
accuracy using these parameterizations.

This interpretation is given further impetus, when we consider the
BaBar data as being two ``experiments'' BaBar1 and BaBar2 --- see
Fig.\ \ref{fig:split-BaBar}.
One sees from Table~3, in terms of $\chi^2_{ndf}$, that the flat-DA
scenario cannot describe {\it both} BaBar ``experiments''
simultaneously with the same accuracy.
\vspace*{2mm}

\begin{tabular}{lr}%
\begin{minipage}{65mm}
%
\hspace*{-5mm}  \includegraphics[width=65mm]{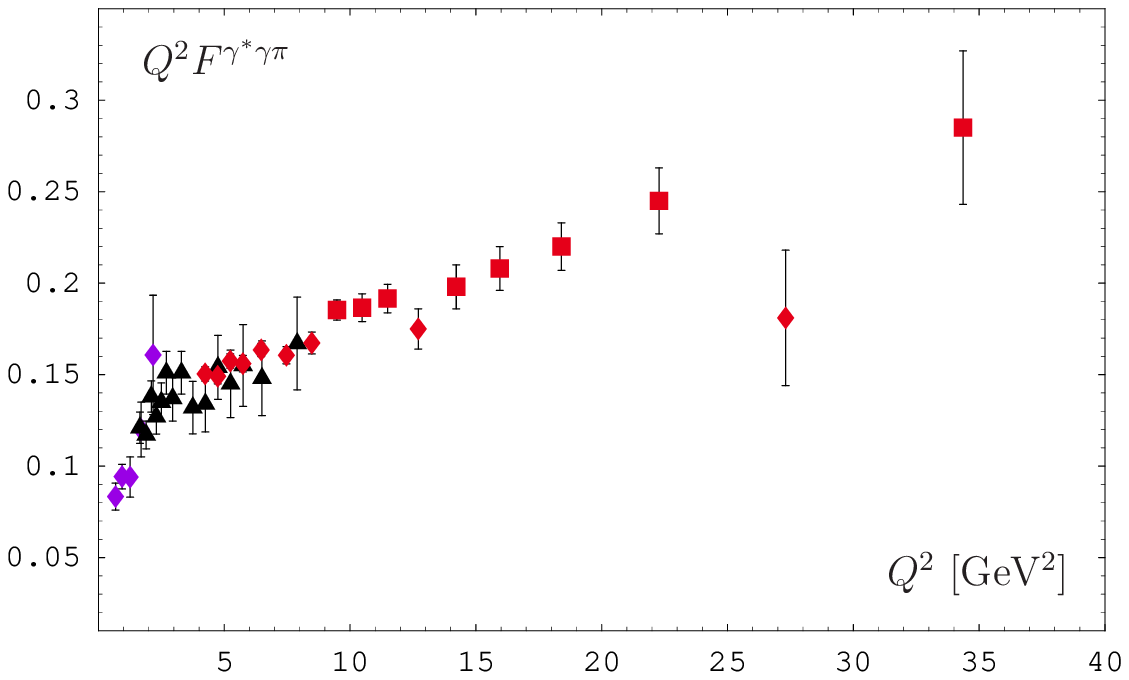}
   \vspace*{-1mm}
{\footnotesize Fig.\,4.~ Split BaBar data:
   \RedTn{\footnotesize \ding{117}} - BaBar1,
   \RedTn{\footnotesize\ding{110}} - BaBar2.}
   \label{fig:split-BaBar}
 \end{minipage}~~& \begin{minipage}{70mm}
\footnotesize{Table 3.~$\chi^2_{ndf}$ for the flat-DA fit. \\
Best-fit values  on the diagonal.}
\vspace*{1mm}

{\begin{tabular}{p{0.15\textwidth} p{0.17\textwidth} p{0.17\textwidth}}
               & \RedTn{\footnotesize \ding{117}} BaBar1
               & \RedTn{\footnotesize\ding{110}}  BaBar2 \\  \hline
 \hspace*{-2mm}BaBar1
               & $3.3$    & $0.33$      \\  \hline
 \hspace*{-2mm}BaBar2$\vphantom{^\big|_|}$
               & \vspace*{-2mm}$3.5$
               & \vspace*{-2mm}$0.26$    \\ \hline
\end{tabular}
\label{table:split-BaBar}
}
\end{minipage}
   \end{tabular}

\section{Conclusions}
\label{sec:concl}

We have studied in detail the pion-photon transition form factor
using light-cone sum rules and including QCD radiative
corrections up to the two loop level.
We also took into account twist-four contributions.
It has been our goal to derive predictions for
$Q^2F^{\gamma\gamma^{*}\pi}$
using several pion distribution amplitudes that can be compared with the
available experimental data.
Our results have been deduced within the convolution scheme of QCD for
distribution amplitudes that vanish at the endpoints $x=0,1$.
They turn out to be unable to match the data of the BaBar Collaboration
for momenta beyond $10$~GeV${}^2$, which grow with increasing $Q^2$.
We analyzed this behavior and argued that proposed scenarios, which make
use of flat pion distribution amplitudes, cannot match the high-$Q^2$
BaBar data and those of the CLEO and the CELLO Collaborations
simultaneously, because the latter demand distribution amplitudes that
vanish at the endpoints.

\section*{Acknowledgments}
This report is dedicated to the 75th birthday of Anatoly Efremov.
We are grateful to A.\ P.\ Bakulev for a fruitful collaboration.
This work was partially supported by the Heisenberg--Landau
Program (Grant 2009) and the Russian Foundation for Fundamental
Research (Grants 07-02-91557 and 09-02-01149).


\end{document}